\documentclass[aps,prl,reprint,twocolumn,superscriptaddress,showpacs,	showkeys,	amsfonts,	amsmath,	amssymb]{revtex4-2}

\usepackage{amsmath}
\usepackage{graphicx}
\usepackage{gensymb}
\graphicspath{{./figure_files/}}
\usepackage{xcolor}
\newcommand{\red}[1]{{\color{red}#1}}

\begin{document}

\title{Measurement of the Charge-Averaged Elastic Lepton-Proton Scattering Cross Section by the OLYMPUS Experiment}
\date{\today}

\author{J.~C.~Bernauer}
\email{jan.bernauer@stonybrook.edu}
\altaffiliation{Now at Stony Brook University, Stony Brook, NY, 11794, USA and Riken BNL Research Center, Upton, NY, 11793, USA}
\affiliation{Massachusetts Institute of Technology, Cambridge, MA 02139, USA}

\author{A.~Schmidt}
\email{axelschmidt@gwu.edu}
\altaffiliation{Now at George Washington University, Washington, DC, 20052, USA}
\affiliation{Massachusetts Institute of Technology, Cambridge, MA 02139, USA}

\author{B.\,S.~Henderson}\affiliation{Massachusetts Institute of
  Technology, Cambridge, MA 02139, USA}

\author{L.D.~Ice}\affiliation{Arizona State University, Tempe, AZ 85281, USA}

\author{D.~Khaneft}\affiliation{Johannes Gutenberg-Universit\"at,
  Mainz, Germany}

\author{C.~O'Connor}\affiliation{Massachusetts Institute of
  Technology, Cambridge, MA 02139, USA}

\author{R.~Russell}\affiliation{Massachusetts Institute of Technology,
  Cambridge, MA 02139, USA}

\author{N.~Akopov}\affiliation{Alikhanyan National Science Laboratory
  (Yerevan Physics Institute), Yerevan, Armenia}

\author{R.~Alarcon}\affiliation{Arizona State University, Tempe, AZ 85287,
  USA}

\author{O.~Ates}
\affiliation{Hampton University, Hampton, VA 23668, USA}

\author{A.~Avetisyan}\affiliation{Alikhanyan National Science
  Laboratory (Yerevan Physics Institute), Yerevan, Armenia}

\author{R.~Beck}\affiliation{Rheinische
  Friedrich-Wilhelms-Universit\"at, Bonn, Germany}

\author{S.~Belostotski}\altaffiliation{deceased}\affiliation{Petersburg Nuclear Physics
  Institute, Gatchina, Russia}

\author{J.~Bessuille}\affiliation{Massachusetts Institute of
  Technology, Cambridge, MA 02139, USA}

\author{F.~Brinker}\affiliation{Deutsches Elektronen-Synchrotron,
  Hamburg, Germany}

\author{J.\,R.~Calarco}\affiliation{University of New Hampshire, Durham,
  NH 03824, USA}

\author{V.~Carassiti}\affiliation{Universit{\`a} degli Studi di Ferrara and Istituto
  Nazionale di Fisica Nucleare sezione di Ferrara, Ferrara, Italy}

\author{E.~Cisbani}\affiliation{Istituto Nazionale di Fisica Nucleare
  sezione di Roma and Istituto Superiore di Sanit\`a, Rome, Italy}

\author{G.~Ciullo}\affiliation{Universit{\`a} degli Studi di Ferrara and Istituto
  Nazionale di Fisica Nucleare sezione di Ferrara, Ferrara, Italy}\

\author{M.~Contalbrigo}\affiliation{Universit{\`a} degli Studi di Ferrara and
  Istituto Nazionale di Fisica Nucleare sezione di Ferrara, Ferrara,
  Italy}

\author{R.~De Leo}\affiliation{Istituto Nazionale di Fisica Nucleare
  sezione di Bari, Bari, Italy}

\author{J.~Diefenbach}
\affiliation{Hampton  University, Hampton, VA 23668, USA}

\author{T.\,W.~Donnelly}\affiliation{Massachusetts Institute of
  Technology, Cambridge, MA 02139, USA}

\author{K.~Dow}\affiliation{Massachusetts Institute of Technology,
  Cambridge, MA 02139, USA}

\author{G.~Elbakian}\affiliation{Alikhanyan National Science
  Laboratory (Yerevan Physics Institute), Yerevan, Armenia}

\author{P.\,D.~Eversheim}\affiliation{Rheinische
  Friedrich-Wilhelms-Universit\"at, Bonn, Germany}

\author{S.~Frullani}\altaffiliation{deceased}\affiliation{Istituto Nazionale di Fisica Nucleare
  sezione di Roma and Istituto Superiore di Sanit\`a, Rome, Italy}

\author{Ch.~Funke}\affiliation{Rheinische
  Friedrich-Wilhelms-Universit\"at, Bonn, Germany}

\author{G.~Gavrilov}\affiliation{Petersburg Nuclear Physics Institute,
  Gatchina, Russia}

\author{B.~Gl\"aser}\affiliation{Johannes Gutenberg-Universit\"at,
  Mainz, Germany}

\author{N.~G\"orrissen}\affiliation{Deutsches Elektronen-Synchrotron,
  Hamburg, Germany}

\author{D.\,K.~Hasell}\affiliation{Massachusetts Institute of Technology,
  Cambridge, MA 02139, USA}

\author{J.~Hauschildt }\affiliation{Deutsches Elektronen-Synchrotron,
  Hamburg, Germany}

\author{Ph.~Hoffmeister}\affiliation{Rheinische
  Friedrich-Wilhelms-Universit\"at, Bonn, Germany}

\author{Y.~Holler}\affiliation{Deutsches Elektronen-Synchrotron,
  Hamburg, Germany}

\author{E.~Ihloff}\affiliation{Massachusetts Institute of Technology,
  Cambridge, MA 02139, USA}

\author{A.~Izotov}\affiliation{Petersburg Nuclear Physics Institute,
  Gatchina, Russia}

\author{R.~Kaiser}\affiliation{University of Glasgow, Glasgow, United
  Kingdom}

\author{G.~Karyan} 
\affiliation{Alikhanyan National
  Science Laboratory (Yerevan Physics Institute), Yerevan, Armenia}
  
\author{J.~Kelsey}\affiliation{Massachusetts Institute of Technology,
  Cambridge, MA 02139, USA}

\author{A.~Kiselev}
\affiliation{Petersburg Nuclear
  Physics Institute, Gatchina, Russia}

\author{P.~Klassen}\affiliation{Rheinische
  Friedrich-Wilhelms-Universit\"at, Bonn, Germany}

\author{A.~Krivshich}\affiliation{Petersburg Nuclear Physics Institute, Gatchina, Russia}

\author{M.~Kohl} \altaffiliation{partially supported by Jefferson Lab}\affiliation{Hampton University, Hampton, VA 23668, USA}

\author{I.~Lehmann}\affiliation{University of Glasgow, Glasgow, United
  Kingdom}

\author{P.~Lenisa}\affiliation{Universit{\`a} degli Studi di Ferrara and Istituto
  Nazionale di Fisica Nucleare sezione di Ferrara, Ferrara, Italy}\

\author{D.~Lenz}\affiliation{Deutsches Elektronen-Synchrotron,
  Hamburg, Germany}

\author{S.~Lumsden}\altaffiliation{deceased}\affiliation{University of Glasgow, Glasgow, United
  Kingdom}

\author{Y.~Ma}
\affiliation{Johannes
  Gutenberg-Universit\"at, Mainz, Germany}

\author{F.~Maas}\affiliation{Johannes Gutenberg-Universit\"at, Mainz,
  Germany}

\author{H.~Marukyan}\affiliation{Alikhanyan National Science
  Laboratory (Yerevan Physics Institute), Yerevan, Armenia}

\author{O.~Miklukho}\affiliation{Petersburg Nuclear Physics Institute,
  Gatchina, Russia}

\author{R.\,G.~Milner}\affiliation{Massachusetts Institute of Technology,
  Cambridge, MA 02139, USA}

\author{A.~Movsisyan} 
  \affiliation{Alikhanyan National Science
  Laboratory (Yerevan Physics Institute), Yerevan, Armenia}
\affiliation{Universit{\`a} degli
  Studi di Ferrara and Istituto Nazionale di Fisica Nucleare sezione
  di Ferrara, Ferrara, Italy}

\author{M.~Murray}\affiliation{University of Glasgow, Glasgow, United
  Kingdom}

\author{Y.~Naryshkin}\affiliation{Petersburg Nuclear Physics
  Institute, Gatchina, Russia}

\author{R.~Perez~Benito}\affiliation{Johannes Gutenberg-Universit\"at,
  Mainz, Germany}

\author{R.~Perrino}\affiliation{Istituto Nazionale di Fisica Nucleare
  sezione di Bari, Bari, Italy}

\author{R.\,P.~Redwine}\affiliation{Massachusetts Institute of
  Technology, Cambridge, MA 02139, USA}

\author{D.~Rodr\'iguez~Pi\~neiro}\affiliation{Johannes Gutenberg-Universit\"at, Mainz, Germany}

\author{G.~Rosner}\affiliation{University of Glasgow, Glasgow, United
  Kingdom}

\author{U.~Schneekloth}\affiliation{Deutsches Elektronen-Synchrotron,
  Hamburg, Germany}

\author{B.~Seitz}\affiliation{University of Glasgow, Glasgow, United
  Kingdom}

\author{M.~Statera}\affiliation{Universit{\`a} degli Studi di Ferrara
  and Istituto Nazionale di Fisica Nucleare sezione di Ferrara,
  Ferrara, Italy}\

\author{A.~Thiel}\affiliation{Rheinische
  Friedrich-Wilhelms-Universit\"at, Bonn, Germany}

\author{H.~Vardanyan}\affiliation{Alikhanyan National Science
  Laboratory (Yerevan Physics Institute), Yerevan, Armenia}\

\author{D.~Veretennikov}\affiliation{Petersburg Nuclear Physics
  Institute, Gatchina, Russia}

\author{C.~Vidal}\affiliation{Massachusetts Institute of Technology,
  Cambridge, MA 02139, USA}

\author{A.~Winnebeck}
\affiliation{Massachusetts Institute of
  Technology, Cambridge, MA 02139, USA}

\author{V.~Yeganov }\affiliation{Alikhanyan National Science
  Laboratory (Yerevan Physics Institute), Yerevan, Armenia}

\collaboration{The OLYMPUS Collaboration}\noaffiliation{}

\begin{abstract}
We report the first measurement of the average of the electron-proton and positron-proton 
elastic scattering cross sections. This lepton charge-averaged cross section is insensitive
to the leading effects of hard two-photon exchange, giving more robust access to the proton's
electromagnetic form factors. The cross section was extracted from data taken by the OLYMPUS
experiment at DESY, in which alternating stored electron and positron beams were scattered from
a windowless gaseous hydrogen target. Elastic scattering events were identified from the 
coincident detection of the scattered lepton and recoil proton in a large-acceptance toroidal
spectrometer. The luminosity was determined from the rates of M\o ller, Bhabha and elastic
scattering in forward electromagnetic calorimeters. The data provide some selectivity between
existing form factor global fits and will provide valuable constraints to future fits. 
\end{abstract}

\maketitle

As the lightest stable composite particle emerging from quantum chromodynamics, the proton
is one of the best testing grounds for our understanding of the strong force. One of the ways
of characterizing the proton's internal quark-gluon structure is through measurements of elastic
electron-proton scattering, from which the proton's electromagnetic form factors, $G_E$ and $G_M$,
can be extracted. These form factors reveal information about how electric charge and current are
distributed within (though this relationship is far from simple, see Ref.\ \cite{Miller:2018ybm}), and
provide a touchstone for the verification of theoretical descriptions and computational approaches.
For large $Q^2$, the progress in precision measurements is hampered by the unresolved discrepancy
between measurements of the proton's elastic form factor ratio, $\mu_p G^p_E / G^p_M$, using polarization techniques~\cite{
      Hu:2006fy, MacLachlan:2006vw, Gayou:2001qt, Punjabi:2005wq,
      Jones:2006kf, Puckett:2010ac, Paolone:2010qc,Puckett:2011xg}, and those  obtained using the traditional
    Rosenbluth technique in unpolarized cross section
    measurements~\cite{Litt:1969my, Bartel:1973rf, Andivahis:1994rq,
      Walker:1993vj, Christy:2004rc, Qattan:2004ht}.
      
    One hypothesis for the cause of this discrepancy is a contribution to the
    cross section from hard two-photon exchange (TPE), which is not included in standard
    radiative corrections and would affect the two measurement techniques
    differently~\cite{ Guichon:2003qm, Blunden:2003sp, Chen:2004tw,
      Afanasev:2005mp, Blunden:2005ew, Kondratyuk:2005kk}.
         
    Standard radiative correction prescriptions account for two-photon exchange
    only in the soft limit, in which one photon carries negligible
    momentum~\cite{Mo:1968cg, Maximon:2000hm}. There is no
    model-independent formalism for calculating hard TPE.  Some
    model-dependent calculations suggest that TPE is responsible for the
    form factor discrepancy~\cite{Chen:2004tw, Afanasev:2005mp,
      Blunden:2005ew, Kondratyuk:2005kk} while others contradict that
    finding~\cite{Bystritskiy:2006ju, Kuraev:2007dn}. The current status
    of the recent experimental and theoretical progress on two-photon 
    exchange is summarized in Ref.~\cite{Afanasev:2017gsk}.
    
    While most models predict negligible effects of hard two-photon exchange on measurements using polarization, such measurements can only extract 
    the form factor ratio. A separation of $G_E$ and $G_M$ requires absolute measurements of the lepton-proton cross sections,
    which are affected by hard TPE. To leading order, TPE effects depend on the charge sign of the lepton. Therefore, a
    charge-averaged cross section is far less sensitive to TPE. We report here on the first precision determination of a
    charge-averaged cross section of $e^{\pm}-p$ scattering.
    
    OLYMPUS's main goal was to measure the ratio of the cross sections for positron-proton and electron-proton scattering,
    a quantity which gives direct access to the two-photon exchange correction. OLYMPUS was optimized for this purpose, and
    the results were published in Ref.~\cite{Henderson:2016dea}. However, careful further analysis allowed us to extract
    charge-averaged cross sections. They cover an interesting kinematical region, where existing form factor fits show a
    turn-over behavior for $G_M$, and where the existing data for $e^- -p$ scattering are somewhat lacking, leading to large
    model uncertainties.
    
    Only a brief overview of the OLYMPUS experiment is given here, and we refer to Ref.\ \citep{Milner:2013daa} for a detailed
    description of the detector. OLYMPUS was the last experiment to take data at the DORIS electron-positron storage ring
    at DESY, Hamburg, Germany. In total, an integrated luminosity of $4.5$~fb$^{-1}$ was collected.
    The 2.01~GeV stored beams with up to 65~mA of current passed through an internal, unpolarized hydrogen gas target
    with an areal density of approximately $3\times10^{15}$~atoms/cm$^2$~\citep{Bernauer:2014pva}.
    The accelerator magnet power supplies were modified to allow the daily change of beam species.
    
    The main detector, a toroidal magnetic spectrometer, was based on the former MIT-Bates BLAST
    detector~\citep{Hasell:2009zza}, with the two horizontal sections instrumented with large acceptance
    ($20\degree<\theta<80\degree$, $-15\degree<\phi<15\degree$) drift chambers (DC) for 3D particle
    tracking and walls of time-of-flight scintillator bars (ToF) for triggering and particle identification.
    The data presented here were collected entirely with positive-tracks-outbending toroid polarity in order to suppress background
    rates in the DC, so that low-energy electrons were bent back to the beam axis and away from the detectors.

    Two new detector systems were designed and built to monitor the
    luminosity. These were symmetric M{\o}ller/Bhabha calorimeters (SYMB)
    at $1.29\degree$~\citep{Benito:2016cmp} and two telescopes of three
    triple gas electron multiplier (GEM) detectors~\citep{Ates:2014aa}
    interleaved with three multi-wire proportional chambers (MWPC)
    mounted at $12\degree$.

    The trigger system selected candidate events that resulted from a
    lepton and proton detected in coincidence in opposite sectors.
    The data were acquired and stored via the CBELSA/TAPS data acquisition system~\citep{Thiel:2012yj}.
    
    The positions of all detector elements were determined via optical surveys and the magnetic
    field was mapped throughout the complete tracking volume with the magnet unmoved from its final position
    in the experiment~\citep{Bernauer:2016cc}.

    Acceptances, radiative corrections and efficiencies were accounted for via a realistic Monte Carlo (MC) simulation.
    The MC parameters, for example beam position and beam current, were adjusted dynamically to match the values recorded by the
    slow control system to simulate the time dependence of these quantities. This approach also rigorously captured
    the possible correlation between parameters. The MC simulation used a radiative event generator
    developed specifically for OLYMPUS~\citep{Russell:2016aa,Schmidt:2016aa}.
    This generator produced lepton-proton events weighted by several different radiative cross
    section models. In this Letter, we present the results following the
    Maximon-Tjon~\citep{Maximon:2000hm} prescription. Higher order radiative
    corrections are taken into account through exponentiation. 

    Particle trajectories and energy losses were simulated using Geant4,
    with custom digitization routines to produce output identical in format to
    actual measured data. This step included efficiency and resolution simulations
    whose parameters were determined from data. Both the simulated and the
    real data were then analyzed with identical software.

    Track reconstruction used a fast hierarchical pattern matching algorithm
    to identify track candidates. Initial track parameters were then determined
    via two distinct track fit algorithms. The design of the drift chambers
    and the running conditions in OLYMPUS led to some track-fitting ambiguities
    that were difficult for the algorithms to resolve. While the algorithms
    did well for most constellations, they failed for certain pathological cases. 
    However, the two algorithms struggled in different cases, so that the combination
    of both algorithms secured the reconstruction with high efficiency over the whole
    phase space. 

    Particle identification was achieved by a combination of track curvature 
    direction, indicating the particle charge, and the correlation between
    momentum and ToF to cleanly separate positrons from protons. 

    The efficiency of the drift chambers was determined by performing track
    reconstruction without considering one of the drift chamber superlayers
    and then considering whether or not hits were present in the ignored
    superlayer. This technique was used to develop highly granular efficiency
    maps of each drift cell. These maps were used directly in the detector
    simulation. While the majority of the drift cells had efficiency $>95\%$,
    several had reduced efficiency, likely because of high discriminator
    thresholds. These inefficient cells had only a small effect on the
    overall tracking efficiency because of the redundancy of the six
    superlayers.

    The efficiency of the time-of-flight scintillators was assessed using
    the combination of cosmic ray studies, data taken with a prescaled efficiency
    trigger, and Geant4 simulation. The efficiency was greater than 99\% for protons
    and greater than 97\% for electrons/positrons. The ToF efficiency model was also implemented
    in the OLYMPUS simulation.

    The track reconstruction efficiency was assessed by selecting elastically recoiling protons
    in one sector and looking for the corresponding scattered lepton in the other sector.
    Within the precision of the study ($\approx 1\%$), there was no indication of inefficiency
    beyond that caused by ToF and drift chamber inefficiencies, and we therefore assign a 1\% 
    normalization uncertainty for any tracking inefficiency. An additional 2\% absolute normalization
    was estimated for other sources not tested by this method, e.g. for the trigger efficiency.
            
    Four independent elastic event selection routines were developed~\citep{Henderson:2016aa,
    Russell:2016aa, Schmidt:2016aa,Bernauer:analysis}, which allowed us to assess the degree of event-selection bias.
    While the four approaches differ in detail, they all exploit the fact that, for a coincidence
    measurement of elastic scattering, the kinematics are overdetermined and that selection cuts
    on the self-consistency of the kinematics can be used to suppress inelastic background. 
    The routine of Ref.~\cite{Henderson:2016aa} used wide cuts in lepton-proton vertex time correlation, vertex-position correlation, 
    polar-angle correlation, and momentum correlation to reduce inelastic background, before estimating any 
    remaining background using sidebands in the azimuthal distribution of track pairs. 
    The routine of Ref.~\cite{Schmidt:2016aa} performed different selection cuts and is distinguished by 
    performing particle-ID at the level of track pairs, rather than individual tracks. 
    The routine of Ref.~\cite{Russell:2016aa} examined background over a two-dimensional space
    of polar and azimuthal angle correlation. 
    The routine of Ref.~\cite{Bernauer:analysis} built an elastic-pair probability for track pairs based
    on their vertex, time and angle correlations, and the missing energy assuming elastic kinematics. 
    Low probability combinations were rejected. The surviving best pair for each event were then used for
    the rate extraction, with a background estimate based on the coplanarity in $Q^2$ slices. 

\begin{figure}[htpb]
      \centering
      \includegraphics[width=\linewidth]{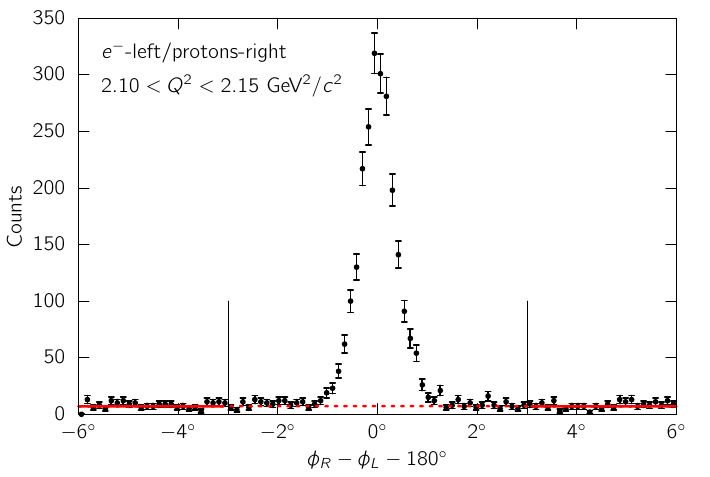}
      \caption{\label{fig:bkg} Background was estimated and subtracted in Ref.~\cite{Schmidt:2016aa} using fits to
        the sidebands of the distribution of the difference in azimuth of lepton ($\phi_L$) and proton ($\phi_R$) track pairs
        after all other elastic event selection criteria were applied. The background was largest at high $Q^2$, as shown
        here, with little difference between $e^-$ and $e^+$ modes.}
    \end{figure}

    The background remaining after elastic event selection was subtracted. 
    The four analyses found similar levels of background for both lepton species and found
    that the background level was higher with increasing $Q^2$. Because their event selection cuts
    differed in tightness, the four analyses varied in the amount of background they subtracted,
    ranging between 5\% and 20\% for the highest $Q^2$ bin. Figure \ref{fig:bkg} shows an example
    of the background fit in one analysis for one of the highest $Q^2$ bins.

    The total recorded data were screened for optimal running conditions, and a subset
    corresponding to $3.1$~fb$^{-1}$ of integrated luminosity (the same subset as in Ref.\ \cite{Henderson:2016dea}) 
    was selected for the results presented here.

    OLYMPUS was optimized for a measurement of the cross section ratio between the two beam species,
    and therefore it employed three independent systems to determine \emph{relative} luminosity:
    from the elastic rate in the two $12\degree$ telescopes, the M{\o}ller-Bhabha rate in the SYMB, and from the
    beam current and target density recorded by the slow control system.
    For an \emph{absolute} measurement of the luminosity, none of the systems is optimal:
    \begin{itemize}
    \item Fundamentally, the $12\degree$ telescopes measure the same process as the main spectrometer
      and can therefore not give an absolute measurement. They could, however, extend the $Q^2$ range of
      the measurement, so that an external determination of the cross section at this smaller value,
      (for example, from a global fit) could establish the normalization and thus a quasiabsolute cross section
      for the remaining data points. However, the $12\degree$ telescope acceptance and absolute
      efficiency are not known well enough to produce a sensible result. 
    \item The slow control system could, in principle, give an absolute normalization. However,
      uncertainties from the target temperature, which affects the density, as well as the absolute
      calibration of the beam current could not be quantified with a reliable error estimate.
    \item The most robust SYMB analysis made use of multi-interaction events, in which a symmetric 
      M\o ller or Bhabha event occurred in the same bunch as an unrelated forward-scattering elastic
      $ep$ event. This method takes advantage of the cancellation of many systematic effects 
      when determining the relative luminosity between beam species. However, these effects do
      not cancel in the determination of the absolute luminosity, resulting in an uncertainty of 7\%. 
      This method is used for normalizing the cross sections reported in this work.
    \end{itemize}
    We note that the results of the SYMB and slow control differ only by about 1\%.
   
    We report the cross section determined from the average of the results of the four independent analyses. 
    We further use the variance between the analyses to estimate systematic uncertainties from event selection choices. 
    However, we first remove the effect of normalization differences between the analyses. We find, for each analysis,
    the normalization factor that minimizes the difference of the analysis to the original average. After renormalization,
    we then determine the remaining variance and use this as an additional point-to-point uncertainty. 
    The stdandard deviation of the normalization constants, 1.5\%, is added as an additional contribution to the global
    normalization uncertainty.
    The systematic difference between cross sections determined from the lepton-left/proton-right versus
    proton-right/lepton-left topologies is used to assess the systematic uncertainty from mis-modeling of the 
    detector acceptance (0.7\%). In total, we achieve a global normalization uncertainty of 7.5\%,
    dominated by the luminosity uncertainty. Table\ \ref{tab:systematics} gives an overview.

    \begin{table}
      \caption{\label{tab:systematics}Contributions to the systematic
        uncertainty in the global normalization}
    \begin{tabular}{l|c}
     \hline\hline
    Source & Uncertainty in the normalization\\
           \hline
    Luminosity & 7.0\%\\
    Efficiency & 2.0\%\\
    Event selection & 1.5\%\\
    Track reconstruction & 1.0\%\\
    Detector acceptance & 0.7\%\\
    Live-time correction & 0.5\%\\
    \hline
    Total & 7.5\% \\
    \hline \hline
    \end{tabular}
    \end{table}

    \begin{figure}
      \includegraphics[width=\linewidth]{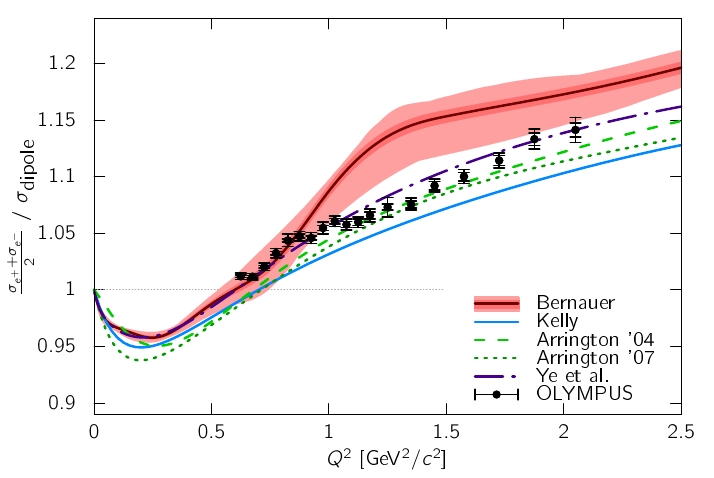}
      \caption{\label{fig:results} The data for the charge-average cross section as a function of
        $Q^2$, in comparison with a series of predictions from form factor fits
        \cite{Bernauer:2013tpr,PhysRevC.70.068202,Arrington:2003qk,Arrington:2007ux,Ye:2017gyb}. The Bernauer \cite{Bernauer:2013tpr}
        prediction is shown with statistical (inner band) and model dependency systematical error
        (added linearly to statistical error, outer band). As can be deduced from the width of the bands
        and the differences between the models, prior data do not strongly constrain models in the range of
        $0.8<Q^2<2.5$~GeV$^2/c^2$; this work can provide some remedy.
      }
    \end{figure}

    The OLYMPUS determination of the charge-average cross section, as a function of
    $\epsilon$ and $Q^2$ is provided in Table\ \ref{tab:results}.
    A comparison of our results with a selection of fits is shown in Fig.~\ref{fig:results}.
    The fits presented here use different methods to minimize the influence of TPE on the
    extracted form factors. All use both Rosenbluth as well as polarized data in their fits,
    and assume that the influence of TPE on the ratio extracted from polarized data is minimal.
    Kelly \cite{PhysRevC.70.068202} omits $G_E$ results for $Q^2>1\ (\mathrm{GeV}/c)^2$ and
    relies on ratio determinations from polarized experiments and $G_M$ values extracted from
    $e^- -p$ scattering, but does not correct them for hard TPE effects. While the effect of TPE
    on the extraction is small compared to the effect on $G_E$ at these $Q^2$, it is not clear
    \emph{a priori} how large the effect is, and how the uncorrected data at smaller $Q^2$ affect
    the high-$Q^2$ behavior. Arrington (2004) \cite{Arrington:2003qk} uses a phenomenological correction
    to the cross sections with a linear dependence in $\epsilon$ and fixed scale of 6\%. Arrington {\em et al.} (2007)
    \cite{Arrington:2007ux} and Ye {\em et al.}\ \cite{Ye:2017gyb} use theoretical TPE calculations and complement them for data points
    $>1\ (\mathrm{GeV}/c)^2$ with an {\em ad hoc} additional effect that is linear in $\epsilon$ and a scaled
    with logarithmic dependence on $Q^2$. Bernauer \cite{Bernauer:2014pva} uses a two-parameter phenomenological
    model, a combination of the Feshbach correction, valid at $Q^2=0$, and a linear model with logarithmic
    scaling in $Q^2$, applied to data at all $Q^2$, fitting form factor parameters and TPE parameters together.

    The data presented here connect the well-constrained region below 1 $(\mathrm{GeV}/c)^2$ with the region
    between 1 and 2 $(\mathrm{GeV}/c)^2$ where TPE effects are more prominent. The fit by Bernauer preferred a
    strong cusp-like structure in $G_M$ around 1.3 $(\mathrm{GeV}/c)^2$, while the other, less flexible, fits,
    have a smoother transition. The data seem to be in better agreement with the latter, but a more detailed
    study of the effects of the new dataset on form factor fits must follow. 
     
    \begin{figure}[htpb]
      \centering
      \includegraphics[width=\linewidth]{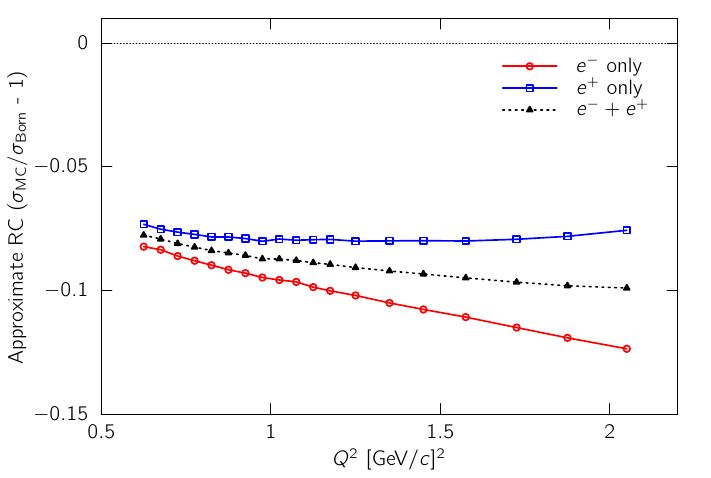}
      \caption{\label{fig:rc} The approximate radiative correction, estimated by taking the 
        ratio of the simulated cross sections with and without the inclusion of radiative effects.
        The charge-odd contribution is a sizeable fraction of the total at high $Q^2$.}
    \end{figure}

    The advantage of the charge-averaging technique is that it suppresses all of the charge-odd radiative corrections.
    The suppression of hard TPE is advantageous because of the uncertainties associated with calculating it, but there
    may be additional benefits as well. Bremsstrahlung from the proton poses a similar challenge to hard TPE since it
    depends on an off-shell proton current. The interference term between electron and proton bremsstrahlung is one of
    the suppressed charge-odd effects, which, combined, grow in magnitude to become a sizable fraction of the total
    correction at higher $Q^2$, shown in Fig.~3. By forming the charge average, the dominant part of the radiative
    correction is from radiation from the electron legs, which is under better theoretical control.

\begin{acknowledgments}
  We thank the DORIS machine group
  and the various DESY groups that made this experiment possible.  We
  gratefully acknowledge the numerous funding agencies: the
  Science Committee of Armenia, Grant 18T-1C180, the Deutsche
  Forschungsgemeinschaft, the European Community-Research
  Infrastructure Activity, the United Kingdom Science and Technology
  Facilities Council and the Scottish Universities Physics Alliance,
  the United States Department of Energy and the National Science
  Foundation, and the Ministry of Education and Science of the Russian
  Federation. R.~G.~M. also acknowledges the generous support of the
  Alexander von Humboldt Foundation, Germany. M.K. is partially supported by Jefferson Lab.
\end{acknowledgments}

\begin{table*}
\caption{\label{tab:results} Cross sections measured by OLYMPUS, using the exponentiated Maximon and Tjon radiative corrections prescription. 
Uncertainties are statistical and point-to-point systematic. There is a further 7.5\% normalization uncertainty that is common to all data points.}
\begin{tabular}{ c c |c | c| c }
\hline
\hline
$\langle Q^2\rangle$ [GeV$^2/c^2$] & $\langle \epsilon \rangle$ & $\sigma_{e^-p}$/std.~dipole & $\sigma_{e^+p}$/std.~dipole & Avg.~$\sigma_{ep}$/std.~dipole \\
\hline
0.624 & 0.898 & $1.0140 \pm 0.0014 \pm 0.0039$ &  $1.0097 \pm 0.0013 \pm 0.0031$ &  $1.0119 \pm 0.0010 \pm 0.0027$ \\
0.674 & 0.887 & $1.0155 \pm 0.0015 \pm 0.0043$ &  $1.0076 \pm 0.0015 \pm 0.0037$ &  $1.0116 \pm 0.0011 \pm 0.0028$ \\
0.724 & 0.876 & $1.0236 \pm 0.0017 \pm 0.0019$ &  $1.0169 \pm 0.0016 \pm 0.0053$ &  $1.0202 \pm 0.0012 \pm 0.0030$ \\
0.774 & 0.865 & $1.0361 \pm 0.0019 \pm 0.0024$ &  $1.0287 \pm 0.0019 \pm 0.0062$ &  $1.0324 \pm 0.0013 \pm 0.0039$ \\
0.824 & 0.853 & $1.0475 \pm 0.0022 \pm 0.0048$ &  $1.0397 \pm 0.0021 \pm 0.0069$ &  $1.0436 \pm 0.0015 \pm 0.0053$ \\
0.874 & 0.841 & $1.0496 \pm 0.0024 \pm 0.0060$ &  $1.0451 \pm 0.0023 \pm 0.0025$ &  $1.0473 \pm 0.0017 \pm 0.0039$ \\
0.924 & 0.829 & $1.0473 \pm 0.0028 \pm 0.0060$ &  $1.0443 \pm 0.0027 \pm 0.0039$ &  $1.0458 \pm 0.0019 \pm 0.0045$ \\
0.974 & 0.816 & $1.0545 \pm 0.0031 \pm 0.0043$ &  $1.0547 \pm 0.0029 \pm 0.0061$ &  $1.0546 \pm 0.0021 \pm 0.0051$ \\
1.024 & 0.803 & $1.0622 \pm 0.0034 \pm 0.0055$ &  $1.0591 \pm 0.0033 \pm 0.0043$ &  $1.0606 \pm 0.0024 \pm 0.0042$ \\
1.074 & 0.789 & $1.0600 \pm 0.0038 \pm 0.0082$ &  $1.0553 \pm 0.0036 \pm 0.0034$ &  $1.0576 \pm 0.0026 \pm 0.0040$ \\
1.124 & 0.775 & $1.0619 \pm 0.0042 \pm 0.0059$ &  $1.0577 \pm 0.0040 \pm 0.0044$ &  $1.0598 \pm 0.0029 \pm 0.0036$ \\
1.174 & 0.761 & $1.0653 \pm 0.0047 \pm 0.0064$ &  $1.0663 \pm 0.0044 \pm 0.0056$ &  $1.0658 \pm 0.0032 \pm 0.0049$ \\
1.246 & 0.739 & $1.0729 \pm 0.0038 \pm 0.0080$ &  $1.0730 \pm 0.0036 \pm 0.0092$ &  $1.0729 \pm 0.0026 \pm 0.0084$ \\
1.347 & 0.708 & $1.0769 \pm 0.0046 \pm 0.0059$ &  $1.0743 \pm 0.0042 \pm 0.0045$ &  $1.0756 \pm 0.0031 \pm 0.0043$ \\
1.447 & 0.676 & $1.0976 \pm 0.0054 \pm 0.0026$ &  $1.0864 \pm 0.0049 \pm 0.0077$ &  $1.0920 \pm 0.0036 \pm 0.0044$ \\
1.568 & 0.635 & $1.0944 \pm 0.0054 \pm 0.0054$ &  $1.1058 \pm 0.0050 \pm 0.0065$ &  $1.1001 \pm 0.0037 \pm 0.0049$ \\
1.718 & 0.581 & $1.1125 \pm 0.0066 \pm 0.0078$ &  $1.1160 \pm 0.0061 \pm 0.0042$ &  $1.1142 \pm 0.0045 \pm 0.0045$ \\
1.868 & 0.524 & $1.1325 \pm 0.0082 \pm 0.0117$ &  $1.1338 \pm 0.0076 \pm 0.0052$ &  $1.1331 \pm 0.0056 \pm 0.0070$ \\
2.038 & 0.456 & $1.1326 \pm 0.0090 \pm 0.0128$ &  $1.1500 \pm 0.0084 \pm 0.0091$ &  $1.1413 \pm 0.0061 \pm 0.0092$ \\
\hline
\hline
\end{tabular}
\end{table*}

\bibliography{references}

\end{document}